\documentclass[aps,prd,onecolumn,groupedaddress,nofootinbib,showpacs, showkeys,amssymb,10pt]{revtex4}
\usepackage[a4paper,left=25mm,right=25mm,top=25mm,bottom=25mm]{geometry}
\usepackage{latexsym,float}
\usepackage{epsfig,graphics}
\usepackage{epstopdf}
\usepackage{amssymb}
\usepackage{amsmath}
\usepackage{verbatim}
\usepackage{multirow}
\usepackage{color}
\usepackage{tikz}
\usepackage[utf8]{inputenc}
\usetikzlibrary{matrix}
\usepackage{float}

\def\gtwid{\mathrel{\raise.3ex\hbox{$>$\kern-.75em\lower1ex\hbox{$\sim$}}}}
\def\ltwid{\mathrel{\raise.3ex\hbox{$<$\kern-.75em\lower1ex\hbox{$\sim$}}}}
\def\square{\kern1pt\vbox{\hrule height 1.2pt\hbox{\vrule width 1.2pt\hskip 3pt
			\vbox{\vskip 6pt}\hskip 3pt\vrule width 0.6pt}\hrule height 0.6pt}\kern1pt}
\begin{document}

\title{Noether symmetry in a nonlocal $f(T)$ Gravity}

\author{Phongpichit Channuie}
\email{channuie@gmail.com}
\affiliation{School of Science, Walailak University, Thasala, Nakhon Si Thammarat, 80160, Thailand}

\author{Davood  Momeni}
\email{davood@squ.edu.om}
\affiliation{Department of Physics, College of Science, Sultan Qaboos University,
	\\P.O. Box 36, P.C. 123, Al-Khodh,, Muscat, Sultanate of Oman}

\begin{abstract}
It is well known that the Noether symmetry approach proves to be very useful not only to fix physically viable cosmological models but also to reduce dynamics and achieve exact solutions. In this work, We examine a formal framework of nonlocal $f(T)$ theory of gravity through the Noether symmetry approach. The Noether equations of the nonlocal $f(T)$ theory are obtained for flat FLRW universe. We analyse the dynamics of the field in a nonlocal $f(T)$ gravity admitted by the Noether symmetry. We observe that there exists a transition from a deceleration phase to the acceleration one in our present analysis. We check statefinder parameters for the obtained solutions which imply that some particular solutions are comparable with the Lambda-CDM model.
\end{abstract}

\date{\today}

\maketitle

%%%%%%%%%%%%%%%%%%%%%%%%%%%
\section{Introduction} 
\label{introduction}
%%%%%%%%%%%%%%%%%%%%%%%%%%%
In addition to the inflationary stage \cite{Inflation} in the early universe, various cosmological observations so far convince that the expansion of the universe is currently accelerating. These experimental results include Type Ia Supernovae \cite{SN}, 
cosmic microwave background (CMB) radiation \cite{Ade:2015xua, Ade:2015lrj, Ade:2014xna, Ade:2015tva, Array:2015xqh, Komatsu:2010fb, Hinshaw:2012aka}, 
large scale structure \cite{LSS}, 
baryon acoustic oscillations (BAO) \cite{Eisenstein:2005su} 
as well as weak lensing \cite{Jain:2003tba}. Regarding the late-time cosmic acceleration, 
there are at least two promising explanations, to date. One of them is to introduce ``dark energy (DE)'' in the context of general relativity. Another attractive approach is to consider the modification of Einstein gravity on the large-scale methodology (for reviews on not only dark energy problem but also modified gravity theories, see e.g., \cite{R-DE-MG}). However, the DE sector remains still unknown.  

As an alternative to standard general relativity, there are possibilities for the theory of such a modification. One of them known as teleparallel gravity is formulated via the Weitzenb\"{o}ck connection \cite{T-G}. 
In teleparallel gravity, the use of torsion (not curvature) is basically implemented contrary to the case of general relativity, in which the Levi-Civita connection is commonly concerned. To be more concrete, the torsion scalar $T$ represents the Lagrangian density of teleparallel gravity. Moreover, the extension of this special case is very similar to that of $f(R)$ gravity \cite{F-R}, where $R$ is the scalar curvature. The resulting theory is named as the $f(T)$ gravity where $f(T)$ is a function of $T$ (for a recent review, see for instance \cite{Cai:2015emx}). Inflationary behavior in the early universe \cite{F-T-Inf} and the late-time cosmic acceleration \cite{F-T-LC} can be promisingly realized 
in $f(T)$ gravity. As a result, various cosmological and astrophysical frameworks based on the $f(T)$ gravity have widely been executed \cite{F(T)-Refs}. Notice that in $f(T)$ gravity, the local Lorentz invariance is broken \cite{L-L-I}, and the relevant investigations on this point have been emerged \cite{RP-LLI}. 

However, when working with the quantum effects, the above machineries could in principle be reconsidered. In Ref.\cite{Deser:2007jk}, there has been considered a way of modifying gravitation, the so-called nonlocal teleparallel  gravity, which is aimed to deal with the quantum effects. It was argued that nonlocal teleparallel formalism plays a powerful tool to study quantum gravitational effects. In order to formulate the unification of inflation in the early universe and the late-time accelerated expansion of the universe, it was shown in Ref.\cite{Nojiri:2007uq} that nonlocal gravity has been modified by adding an $f(R)$ term. In addition, a possible solution for the cosmological constant problem through the nonlocal property of gravitation~\cite{ArkaniHamed:2002fu} 
has been proposed. Moreover, a physical mechanism by which a cosmological constant is screened in the framework of nonlocal gravity has been investigated in Refs.\cite{Nojiri:2010pw,Bamba:2012ky,Zhang:2011uv}. 
Some indications imply that there exist the issue of ghosts \cite{Nojiri:2010pw} in the nonlocal gravity. Various aspects of nonlocal gravity have widely been advertised \cite{NL-Ref}. For a recent review on nonlocal gravity, see e.g., Ref.\cite{Maggiore:2016gpx}. It is worth noting that the action principle for the theory proposed by Maggiore {\it et.\,al.} was first derived in Ref.\cite{Modesto:2013jea}. 

More recently, the nonlocal deformations of teleparallel gravity 
has been analyzed in Ref.~\cite{Bahamonde:2017bps}. 
This theory is called the nonlocal $f(T)$ gravity, which 
can be considered as an extension of nonlocal general relativity to the Weitzenb\"{o}ck spacetime. It has been discussed that there is a possibility to distinguish teleparallel gravity from general relativity with future experiments by detecting the nonlocal effects. The purpose of the present study is to analyse the dynamics of the field in the a nonlocal $f(T)$ Gravity
through the Noether symmetry technique. It has been proven that this approach proved to be very useful not only to fix physically viable cosmological models
with respect to the conserved quantities, e.g., potentials, but also to reduce dynamics and achieve exact solutions. The Noether symmetry technique has been employed to various cosmological scenarios, see e.g. \cite{cap3}-\cite{Kaewkhao:2017evn}.

%%% Organization %%%
This paper is organized as follows: We will start by making a short recap of a formal framework of nonlocal $f(T)$ theory of gravity in Sec.\ref{s2}. Here we derive the relevant equations of motion (EoM) for underlying theory. A review of the Noether symmetry approach will be dictated in Sec.\ref{s3}. In Sec.\ref{s4}, we analyse the dynamics of the field in the a nonlocal $f(T)$ gravity through the Noether sysmetry technique and determine the solutions for the symmetry. In Sec.\ref{s5}, we check statefinder parameters for the obtained solutions. Finally, we conclude our findings in the last section.

%%%%%%%%%%%%%%%%%%%%%%
\section{Formal framework of nonlocal $f(T)$ gravity}
\label{s2}
Let us first develop the formalism of nonlocal modified gravity with torsion $T$. We assume that total action for gravity and matter can be written in the following mainly inspired from $f(R)$ gravity, 
\begin{eqnarray}
&&S=\frac{1}{2\kappa}\int d^4x eT\Big(f(\Box^{-1}T)-1\Big)+\int d^4x e \mathcal{L}_m\label{action}
\end{eqnarray}
Here the gravitational coupling is 
 $\kappa=8\pi G$, speed of light is set $c=1$, $G$ is Newtonian gravitational constant and the local operator $\Box $ is called  d'Alembert operator is defined as  $\Box=e^{-1}\partial_
{\alpha}(e\partial^{\alpha})$ ,for matter contents we write  $\mathcal{L}_m$. Furthermore,  Greek alphabets $\mu'\nu,...$ run from $0...3$ and $\Box^{-1}$ is considered as integral over entirely spacetime manifold. Instead of common factor $\sqrt{-{\rm det}g_{\mu\nu}}$ in GR, we use an equivalent measure  $e=\sqrt{-g}={\rm det}(e_{a}^{\mu})=\sqrt{-{\rm det}(g_{\alpha\beta})}$,  and $T$ is torsion scalar 
 defined by
\begin{equation}\nonumber
T=S^{\:\:\:\mu \nu}_{\rho} T_{\:\:\:\mu \nu}^{\rho}\,,
\end{equation}
and the components of the torsion tensor
$$
T_{\:\:\:\mu \nu}^{\rho}=e_i^{\rho}(\partial_{\mu}
e^i_{\nu}-\partial_{\nu} e^i_{\mu})\,,
$$
$$
S^{\:\:\:\mu \nu}_{\rho}=\frac{1}{2}(K^{\mu
\nu}_{\:\:\:\:\:\rho}+\delta^{\mu}_{\rho} T^{\theta
\nu}_{\:\:\:\theta}-\delta^{\nu}_{\rho} T^{\theta
\mu}_{\:\:\:\theta})\,,
$$
and also for contorsion tensor
$$
K^{\mu \nu}_{\:\:\:\:\:\rho}=-\frac{1}{2}(T^{\mu
\nu}_{\:\:\:\:\:\rho}-T^{\nu \mu}_{\:\:\:\:\:\rho}-T^{\:\:\:\mu
\nu}_{\rho})\,.
$$
Following the methods in non local $f(R)$ theory, it is illustrative to find a scalar-tensor reduction 
for action (\ref{action}) using a pair of auxiliary (commonly accepted as  unphysical fields) fields $\phi=\frac{1}{\Box}T$ and $\xi=-\frac{1}{\Box}(f'(\phi)T)$, the new form for the reduced action is written as follows:
\begin{eqnarray}
&&S=\frac{1}{2\kappa}\int d^4x e\Big[T\Big(f(\phi)-1\Big)-\partial_{\mu}\xi\partial^{\mu}\phi-\xi T
\Big]
+\int d^4x e \mathcal{L}_m\label{action2}
\end{eqnarray}
Note that in (\ref{action2}) the action function $f(\phi)$ is supposed to have any desired form. The form of equations of motion presented in  \cite{Bahamonde:2017bps} reads:
\begin{eqnarray}
2(1-f(\phi)+\xi)\left[ e^{-1}\partial_\mu (e S_{a}{}^{\mu\beta})-E_{a}^{\lambda}T^{\rho}{}_{\mu\lambda}S_{\rho}{}^{\beta\mu}-\frac{1}{4}E^{\beta}_{a}T\right]\nonumber\\
-\frac{1}{2}\Big[(\partial^{\lambda}\xi)(\partial_{\lambda}\phi)E_{a}^{\beta}-(\partial^{\beta}\xi)(\partial_{a}\phi)-(\partial_{a}\xi)(\partial^{\beta}\phi)\Big] -2\partial_{\mu}(\xi-f(\phi))E^\rho_a S_{\rho}{}^{\mu\nu}=  \kappa\Theta^\beta_a\,. \label{2}
\end{eqnarray}
Here $\Theta^\beta_a=e^{a}_{\mu}\mathcal{T}_{a\nu}$ denotes
the energy-momentum tensor for matter field's Lagrangian $\mathcal{L}_m$.
The authors performed cosmological data analysis on a suitable chosen function $f(\phi)=A \exp(n\phi)$. However, in our paper we will below fix $f(\phi)$ using the Noether symmetry approach.

It is worth noting that nonlocal theories for gravity are frequently represented in terms of a set of auxiliary fields (see for example Ref.\cite{DeFelice:2014kma}). Here a general class of non local theories is formulated on the Riemanninan geometry where the Lagrangian of the gravitational sector is defined as a general smooth function of powers of the inverse d’Alembertian operator acting on the Ricci scalar. In such GR extensions, a valid localization procedure is to  introduce auxiliary scalar fields. In order to quantify whether they are representing ghosts or not, one must count the numbers of the degrees of freedom of the localized form of the Lagrangian. Furthermore it is needed to check the equivalence between the dynamics of the local and the original nonlocal action in a specific background. By studying the differential forms of the auxiliary fields, one will find a set of algebraic constraints that should be take into the account in writing the local form in a prescribed frame. In GR, it is very straightforward to pass from the conformal frame to the Einstein frame and find the equivalent potential terms of the action. When we have only linear Ricci scalar term $R$, one can show that may or may not the theory ghost-free dependence. However, the situation highly depends on the choice of the parameters. Consequently in GR, except for the special linear case, this class of nonlocal gravity mainly suffers from the presence of a ghost.

In our nonlocal teleparallel theory, we will have the same argument. As far as our nonlocal action made by linear scalar torsion $T$, the theory is ghost free. The reason is that Einstein-Hilbert action (GR) is dynamically equivalent to the teleparallel gravity at level of action as well as equations of motion \cite{Hayashi}. In our paper, using the Noether symmetry approach we obtained the nonlinear class of the solutions for the general action given in Eq.(\ref{action}). In addition, this action made only by linear torsion term $T$. Consequently the model under study is ghost free.

%%%%%%%%%%%%%%%%%%%%%%%%
\section{A Review of Noether symmetry}
\label{s3}
%%%%%%%%%%%%%%%%%%%%%%%%%
Noether symmetry is defined in the context of dynamical systems. We are mainly interested in studying causal systems where the Lagrangian of the whole dynamical system $L\equiv L(q_i, \dot{q}_i;t),\, 1\leq  i\leq N$,\, is a quadratic function of $\dot{q}_i$ and an arbitrary function of time and configuration coordinates $q_i$. It is mostly acceptable to keep Lagrangian up to the higher orders $\mathcal{O}(\ddot{q}_i)$. It is adequate to define a set of adjoint conjugate momenta 
 $\dot{p}_i-\frac{\partial L}{\partial q_{i}}=0,\ \ p_{i}\equiv\frac{\partial L}{\partial \dot{q}_i}$. 
 What is called the Noether symmetry is the existence of a 
 vector (non unique), Noether vector $\vec{X}$ \cite{cap4,noether3,noether4,noether2,cap3} such that:
\begin{equation}\label{17}
 X=\Sigma_{i=1}^{N}\alpha^i(q)\frac{\partial}{\partial q^i}+
 \dot{\alpha}^i(q)\frac{\partial}{\partial\dot{q}^i}\,{,}
 \end{equation}
 If we can adjust a set of functions 
   $\alpha_{i}(q_j)$, in a such manner  that 
 the Lie derivative of Lagrangian vanishes globally 
(the tangent space of
configurations $T{\cal Q}\equiv\{q_i, \dot{q}_i\}$):
 \begin{equation}\label{19}
 L_X{\cal L}=0\,
 \end{equation}
 It is illustrative to rewrite it as following 
 \begin{equation}\label{18}
 L_X{\cal L}=X{\cal L}=\Sigma_{i=1}^{N}\alpha^i(q)\frac{\partial {\cal L}}{\partial q^i}+
 \dot{\alpha}^i(q)\frac{\partial {\cal L}}{\partial\dot{q}^i}\,{.}
 \end{equation}
 It is easy to show that 
for any existed  $\vec{X}$ 
, the total phase flux  enclosed in a region of space, is conserved along $X$. In fact, it is instructive to show that just by applying Euler-Lagrange equations,
 \begin{equation}\label{20}
 \frac{d}{dt}\frac{\partial {\cal L}}{\partial\dot{q}^i}-
 \frac{\partial {\cal L}}{\partial q^i}=0\,{,}\ \ 1\leq i \leq N.
 \end{equation}
As a result  we  obtain
 \begin{equation}\label{21}
 \Sigma_{i=1}^{N}\frac{d}{dt}\left(\alpha^i\frac{\partial {\cal
 L}}{\partial\dot{q}^i}\right)=L_X{\cal L}\,{.}
 \end{equation}
 We will have a polynomial of variables $(q_i,\dot{q}_j)$ and 
if we can find $\alpha_{i}$  by vanishing the coefficents of all powers of $\dot{q}^i$, then we will show that there exist a 
local conserved charge as the following:
 \begin{equation}\label{22}
Q=\Sigma_{i=1}^{N}\alpha^ip_i
 \end{equation}
 
It has been so far shown that the Noether symmetry strategy is a powerful tool to investigate cosmological large scale dynamics as well as it provides a technique to obtain the exact solutions in various scenarios \cite{cap3}-\cite{Kaewkhao:2017evn}. It is worth noting that the existence of the nonlocal form factors in the nonlocal extensions of the GR is an important issue mainly when one considers the nonperturbative spectrum of the theory in the ultraviolet regimes where the propagator of graviton needs to be corrected appropriately. In the GR when adding nonlocal terms to the classical action in the weak field regime, a.k.a. the coupling constants are considered very small, at the perturbative level it has been demonstrated that the degrees of freedom (dofs) of nonlocal theory are equal to those of the corresponding classical partner, i.e. the Einstein-Hilbert theory. 

However, the perturbative computations should be performed in the vicinity of small deformations of an arbitrary but maximally symmetric spacetime, e.g. de Sitter. Since the background metric is considered as a non flat manifold, still it is very difficult to extend the momentum space to such non flat backgrounds. Note that we need to work in momentum space to make graviton propagators healthy. In non flat, the Fourier mode decomposition is no longer valid because of the absence of a unique vacua. In our nonlocal teleparallel with linear scalar torsion term $T$, the same analysis can be done to show that there should not have any extra dofs more than the ones we expect from GR action. However, in nonperturbative approach the problem drastically changes and in the GR a numbers of dofs differ from a nonlocal theory. For example in a recent paper, authors showed that $(\rm dofs)_{\rm nonloca,dim=4}=8$ instead of the expected value from GR \cite{Calcagni:2018gke}. In our study we can say that in the perturbative level still the dofs remain the same as of the teleparallel gravity. If we impose more symmetries, like Noether symmetry which investigated in this paper, it may help to improve the form factor problem. However, we did not figure out any reference on how to perform and treat field theory in such Weitzenbock space.

%%%%%%%%%%%%%%%%%%%%%%
\section{Noether symmetry equations for nonlocal $f(T)$ theory }\label{s4}
Let us suppose that the non singular, physical metric of spacetime is characterized by a Friedman-Lemaitre-Robertson-Walker (FLRW) metric given by $ds^2=dt^2-a(t)^2(dx^bdx_b)$, where $b=1,2,3$ is spatial coordinate and $a(t) $ is a scale factor. It measures expansion of the whole cosmological Universe as well as its acceleration/deceleration phase. The correspondingly suitable, diagonal tetrad (vierbein) basis is given by $e^{a}_{\mu}={\rm diag}\Big(1,a(t),a(t),a(t)\Big)$. The set of the FLRW equations of Eq.(\ref{action2}) can be obtained as follows:
\begin{eqnarray}
3H^2(1+\xi-f(\phi))&=&\frac{1}{2}\dot{\phi}\dot{\xi}+\kappa(\rho_m+\rho_{\Lambda}+\rho_r),\label{eq1}\\
(2\dot{H}+3H^2)(1+\xi-f(\phi))&=&-\frac{1}{2}\dot{\phi}\dot{\xi}+2H(\dot{\xi}-\dot{f}(\phi))-\kappa(p_{\Lambda}+p_r).\label{eq2}
\end{eqnarray}
Adding the above equations together, we find
\begin{eqnarray}
2H(\dot{\xi}-\dot{f}(\phi))=(2\dot{H}+6H^2)(1+\xi-f(\phi))-\kappa \rho_{m}.\label{eq2t}
\end{eqnarray}
The point-like Lagrangian for the action (\ref{action}) in the FLRW background in configuration spaces $(a,\phi,\xi)$, with matter Lagrangian $\mathcal{L}_m=\rho$, takes the form:
\begin{eqnarray}
&&L(a,\phi,\xi,\dot{a},\dot{\phi},\dot{\xi})=\frac{1}{2\kappa}\Big(-6\dot{a}^2a(f(\phi)-1)-\dot{\phi}\dot{\xi}a^3+
6\dot{a}^2a\xi
\Big)+\rho a^3.
\label{lag}
\end{eqnarray}
We obtain from Eq.(\ref{lag}) the Euler-Lagrang (EL) equations:
\begin{eqnarray}
\ddot{\phi} +3H\dot{\phi} +6H^{2} &=& 0,\label{es16}
\\\ddot{\xi} +3H\dot{\xi} -6H^{2}f'(\phi) &=& 0,\label{es17}
\\ \frac{\ddot{a}}{a}\Big(1+\xi-f(\phi)\Big)+(\dot{H}+5H^2)(1+\xi-f(\phi))-\kappa \rho_{m} &=& 0,\label{es170}
\end{eqnarray}
where we have used Eq.(\ref{eq2t}) to obtain Eq.(\ref{es170}) and a prime denotes derivative with respect to $\phi$. Therefore, the  infinitesimal generator of the Noether symmetry is
\begin{eqnarray}
&&\vec{X}=\alpha\frac{\partial}{\partial a}+\beta \frac{\partial}{\partial \phi}+\gamma \frac{\partial}{\partial \xi}+\dot{\alpha}\frac{\partial}{\partial \dot a}+\dot{\beta} \frac{\partial}{\partial \dot{\phi}}+\dot{\gamma} \frac{\partial}{\partial \dot{\xi}}.
\end{eqnarray}
Here $\alpha,\beta,\gamma$ are in general functions of $\{a,\phi,\xi\}$, and $\dot f\equiv \dot{a}\frac{\partial f}{\partial a}+\dot{\phi} \frac{\partial f}{\partial \phi}+\dot{\xi} \frac{\partial f}{\partial \xi}$. 
The Lie derivative of a given Lagrangian $L$ vanishes $\vec{X}L=0$, provides us the following system of the differential equations:
\begin{eqnarray}
\left(\alpha+2a
\frac{\partial \alpha}{\partial a}\right)\Big(1+\xi-f(\phi)\Big)+a\Big(\gamma-\beta f'(\phi)\Big)&=&0, \label{es1}
\\a^2\frac{\partial \gamma}{\partial a}-12\frac{\partial \alpha}{\partial\phi}\Big(1+\xi-f(\phi)\Big)&=&0,\label{es2}
\\a^2\frac{\partial \beta}{\partial a}-12\frac{\partial \alpha}{\partial\xi}\Big(1+\xi-f(\phi)\Big)&=&0,\label{es3}
\\3\alpha+a\left(\frac{\partial \gamma}{\partial \xi}+\frac{\partial \beta}{\partial \phi}
\right)&=&0,\label{es4}
\\\frac{\partial \gamma}{\partial \phi}&=&0,\label{es5}
\\\frac{\partial \beta}{\partial \xi}&=&0,\label{es6}
\\ a\rho'+3\rho&=&0,\label{es7}
\end{eqnarray}
where prime denotes differentiation with respect to the scale factor $a$. We obtain from Eq.(\ref{es5}) and Eq.(\ref{es6}) 
\begin{eqnarray}
\gamma = \gamma(a,\xi),\quad \beta = \beta(a,\phi),
\end{eqnarray}
and from Eq.(\ref{es7})
\begin{eqnarray}
\rho = \rho_{0}a^{-3},
\end{eqnarray}
with $\rho_{0}$ being an integration constant. What we are going to do next is to solve the system of Eqs.(\ref{es1})-(\ref{es4}). For concreteness, we consider below the solutions in three cases. Other cosmological backgrounds, e.g., the anisotropic Bianchi models or inhomogeneous ones, can be considered as possible works for further studies. Very recently we showed that in a class of higher order theories, it is possible to study Bianchi models using the techniques of Noether symmetry \cite{Channuie:2018now}. There are other works where the authors studied Noether symmetry beyond the FLRW cosmology for example \cite{Jamil:2012zm}. Although working with anisotropic backgrounds makes difficulties to define the point like Lagrangian in a suitable phase space coordinates and to define the associated tangent space to the configuration coordinates, or to define a good average scale factor or effective Hubble parameter, still the problem deserves to be worth studying.

%%%%%%%%%%%%%%%%%%%%%%%%%%%%%%%%%%
\subsection{Solutions for Class A}
%%%%%%%%%%%%%%%%%%%%%%%%%%%%%%%%%%
We start sorting out the solutions by considering Eq.(\ref{es2}-\ref{es4}). We find first particular solutions as follows:
\begin{eqnarray}
\alpha(a) &=& -\frac{c_{1}a}{3}+c_{2},\label{es8}
\\ \gamma (\xi) &=& c_{1}\xi + c_{3},\label{es9}
\\\beta &=& c_{4},\label{es10}
\\ f(\phi) &=& c_{5} e^{\frac{c_{1} \phi }{c_4}} +c_{6},\label{es11}
\end{eqnarray}
where $c_{i}$s are constants. The corresponding Noether conserved charge for these particular solutions is defined by
\begin{eqnarray}
Q=\alpha\frac{\partial L}{\partial \dot a}+\beta \frac{\partial L}{\partial \dot\phi}+\gamma \frac{\partial L}{\partial \dot\xi}\,.
\label{Q1}
\end{eqnarray}
Using solutions given in Eqs.(\ref{es8}-\ref{es11}), we obtain for Class A solutions:
\begin{eqnarray}
Q_{1}=\frac{1}{\kappa}\Big(2\dot{a}a(-c_{1}a+3c_{2})(1-c_{5} e^{\frac{c_{1} \phi }{c_4}} -c_{6}+\xi)- \frac{c_{4}}{2} (\dot{\xi}a^3)-\frac{1}{2}(c_{1}\xi + c_{3}) (\dot{\phi}a^3)\Big).
\label{Q2}
\end{eqnarray}
Note that $f(\phi)$ given in Eq. (\ref{es11}) for $c_6=0,\frac{c_1}{c_4}=n,c_5=A$ can be specified by currently observational data given by SNe Ia + BAO + CC + H0 observations. Here we can constrain $(A,n)$ as $A = -0.009713^{+0.017}_ {-0.021},n = 0.02086^{+0.0013}_{-0.0208}$ \cite{Bahamonde:2017bps}. Unfortunately, it is rather difficult to verify the general solutions for $\{a,\phi,\xi\}$ in the first case.

However, with the de Sitter case, i.e. $H=H_{0}$, the particular solutions can be explicitly obtained. In this circumstance, we find for $f(\phi) = c_{5} e^{\frac{c_{1} \phi }{c_4}} +c_{6} \equiv A\, e^{n\phi} +c_{6}$:    
\begin{eqnarray}
\ddot{\phi} +3H_{0}\dot{\phi} +6H^{2}_{0} &=& 0,\label{es16b101}
\\\ddot{\xi} +3H_{0}\dot{\xi} -6nAH^{2}_{0}e^{n\phi} &=& 0,\label{es17b21}
\end{eqnarray}
where the constrained values of $A$ and $n$ are given above. We simply  obtain the solution for Eq.(\ref{es16b1}) as the following:
\begin{eqnarray}
\phi(t) = -\frac{\kappa_1 e^{-3H_{0} t}}{3H_{0}}-2H_{0} t+\kappa_2,\label{es16b111}
\end{eqnarray}
where $\kappa_{i}$s are integration constants. When inserting $\phi(t)$ into Eq.(\ref{es17b21}), we come up with the second order linear nonhomogeneous differential equations. We use the standard method to solve and the solution reads
\begin{eqnarray}
\xi(t) = \xi_{h}(t) + \xi_{i}(t),\label{es16b12}
\end{eqnarray}
where $\xi_{h}$ is a general solution of the corresponding
homogeneous equation given by
\begin{eqnarray}
\xi_{h}(t) = p_2-\frac{p_1 e^{-3 H_{0} t}}{3H_{0}},\label{esc}
\end{eqnarray}
where $p_{i}$s are integration constants. To specify $\xi_{i}(t)$ that satisfies the nonhomogeneous equation, the solution can be written in the following form: 
\begin{eqnarray}
\xi_{i}(t) = F\left(e^{n\phi(t)}\right),\label{escn}
\end{eqnarray}
where $F$ is a function depends on $e^{n\phi(t)}$. However, we will perform numerical calculations in Sec.(\ref{s6}) in order to figure out the general solutions in this class.
%%%%%%%%%%%%%%%%%%%%%%%%%%%%%%%%
\subsection{Solutions for Class B}
%%%%%%%%%%%%%%%%%%%%%%%%%%%%%%%%
Yet we also find another class of solutions (B) given by
\begin{eqnarray}
\alpha &=& 0,\label{es12}
\\ \gamma &=& -c_{1}\,\,{\rm or}\,\,c_{1},\label{es13}
\\ \beta &=& c_{2}\,\,{\rm or}\,\,-c_{2},\label{es14}
\end{eqnarray}
Notice that with $\gamma$ and $\beta$ obtained above we end up with the same form of $f(\phi)$ as
\begin{eqnarray}
f(\phi) = -\frac{c_1}{c_2}\phi+c_{3},\label{es15}
\end{eqnarray}
Using solutions obtained in Eq.(\ref{es12}-\ref{es15}) we find for Class B solutions:
\begin{eqnarray}
&&Q_{2}= - \frac{c_{2}}{2\kappa} (\dot{\xi}a^3)-\frac{c_{1}}{2\kappa}(\dot{\phi}a^3).
\label{Q3}
\end{eqnarray}
In this class, it seems like the exact solutions can basically obtained. Below we will separately discuss two choices of constants, $(c_{1}\& c_{2})$:
\begin{itemize}
\item (i) $c_{1}=c_{2}=1$

We quantify this first case by adding Eq.(\ref{es16}) to Eq.(\ref{es17}). Thus we find for $f(\phi) = -c_{1}\phi/c_{2} + c_{3}$ with $c_{1}=c_{2}=1$:
\begin{eqnarray}
\dot{\phi} + \dot{\xi}= \frac{\eta_{0}}{a(t)^{3}}.
\label{es19}
\end{eqnarray}
For Eq.(\ref{Q3}) we also obtain
\begin{eqnarray}
\phi + \xi= -2\kappa Q_{2}\int\frac{dt}{a(t)^{3}}.
\label{es20}
\end{eqnarray}
Notice that it is not trivial to determine the general exact solutions. However, the problem can be simply relaxed when we focus on de Sitter case, i.e. $H=H_{0}$. Regarding this special situation, we end up with
\begin{eqnarray}
\ddot{\phi} +3H_{0}\dot{\phi} +6H^{2}_{0} &=& 0,\label{es16b}
\\\ddot{\xi} +3H_{0}\dot{\xi} +6H^{2}_{0} &=& 0,\label{es17b}
\end{eqnarray}
whose solutions read
\begin{eqnarray}
\phi (t) &=& -\frac{k_1 e^{-3H_{0} t}}{3H_{0}}-2H_{0} t+k_2,\label{es16b1}
\\\xi (t) &=& -\frac{k_3 e^{-3H_{0} t}}{3H_{0}}-2H_{0} t+k_4,\label{es17b2}
\end{eqnarray}
where $k_{i}$s are integration constants.

\item (ii) $c_{1}=-c_{2}=1$

However, when $c_{1}=-c_{2}=1$, we instead find by subtracting Eq.(\ref{es16}) to Eq.\ref{es17})
\begin{eqnarray}
\dot{\phi} - \dot{\xi}= \frac{\eta_{0}}{a(t)^{3}},
\label{es21}
\end{eqnarray}
and from Eq.(\ref{Q3})
\begin{eqnarray}
\phi - \xi= 2\kappa Q_{2}\int\frac{1}{a(t)^{3}}dt.
\label{es22}
\end{eqnarray}
In the same situation, it is rather difficult to determine the general exact solutions in this case. We simply relax by considering the de Sitter case, i.e. $H=H_{0}$. Regarding this special case, we obtain
\begin{eqnarray}
\ddot{\phi} +3H_{0}\dot{\phi} +6H^{2}_{0} &=& 0,\label{es16b1t}
\\\ddot{\xi} +3H_{0}\dot{\xi} -6H^{2}_{0} &=& 0,\label{es17b2t}
\end{eqnarray}
whose solutions read
\begin{eqnarray}
\phi (t) &=& -\frac{g_1 e^{-3H_{0} t}}{3H_{0}}-2H_{0} t+g_2,\label{es16b11}
\\\xi (t) &=& -\frac{g_3 e^{-3H_{0} t}}{3H_{0}}+2H_{0} t+g_4,\label{es17b22}
\end{eqnarray}
where $g_{i}$ are integration constants.
\end{itemize}

Clearly, the evolution of the solutions obtained from two classes above are very similar. We continue their verification by performing the numerical calculation given in Sec.(\ref{s6}). Note here that our results for this class coincides with those found in Ref.\cite{Bahamonde:2017sdo}.
%%%%%%%%%%%%%%%%%%%%%%%%%%%%%%%%%%
\subsection{Solutions for Class C}
%%%%%%%%%%%%%%%%%%%%%%%%%%%%%%%%%%
From Eq.(\ref{es2}-\ref{es4}), we find another set of particular solutions (C3):
\begin{eqnarray}
\alpha &=& 0,\label{es83}
\\ \gamma &=& 0,\label{es93}
\\\beta &=& c_{1},\label{es103}
\\ f(\phi) &=& c_{2},\label{es113}
\end{eqnarray}
where $c_{i}$s are constants. The corresponding Noether conserved charge is given by
\begin{eqnarray}
Q_{3}=-\frac{c_{1}}{2\kappa}\dot{\xi}a^{3}.\label{Q13}
\end{eqnarray}
Integrating (\ref{Q13}) we obtain
\begin{eqnarray}
\xi=-\frac{2\kappa Q_{3}}{c_{1}}\int_{0}^{t}\frac{dt'}{a(t')^3}\label{xisol3}
\end{eqnarray}
Let us to focus on EL equations for Lagnranigan (\ref{lag}). We should solve the following system of differential eqs:
\begin{eqnarray}
\ddot{\phi} +3H\dot{\phi} +6H^{2} &=& 0,\label{es160}
\\ (\dot{H}+3H^2)(1+\xi-c_{2})-\kappa \rho_{0} a^{-3}&=& 0,\label{es1701}
\end{eqnarray}
We make derivative from (\ref{es1701}) using (\ref{xisol3}) we obtain
\begin{eqnarray}
\frac{4 Q_3}{c_1 \rho_0}=\frac{a(t) \left(a(t)^2 \dddot{a}(t)+2
   \dot{a}(t)^3+6 a(t) \dot{a}(t) \ddot{a}(t)\right)}{\left(a(t) \ddot{a}(t)+2 \dot{a}(t)^2\right)^2}
\end{eqnarray}
It is quite hard to find exact solution for above equation, if we just focus on de Sitter case $H=H_0$, we obtain the following exsct solutions for fields $\phi,\xi$,
\begin{eqnarray}
&&\phi(t)=-\frac{C_1 e^{-3 H_0 t}}{3 H_0}+C_2-2 H_0 t\\&&
\xi(t)=\frac{2\kappa Q_{3}}{3C_{1} H_0}e^{-3 H_0 t},
\end{eqnarray}
with $C_1=-\frac{12 H_0^2Q_3}{\rho_0}$. We also verify the general solution by performing the numerical computation given in Sec.(\ref{s6}).

%%%%%%%%%%%%%%%%%%%%5
\section{Statefinder Parameters}
\label{s5}
  In connection to DE dynamics, it is adequate to use the deceleration parameter $q$ is is defined by,
\begin{equation}\label{state8.2}
q=-\frac{\ddot{a}}{aH^{2}}=-1-\frac{\dot{H}}{H^{2}}.
\end{equation}
Note that in a general type of modified theory of gravity it is always possible to write the FLRW EoMs in the following effective forms:
\begin{eqnarray}
&&H^2=\frac{\kappa}{3}\sum\rho_i,\ \ \dot{H}=-\frac{\kappa}{2}\sum p_i.
\end{eqnarray}
As a result, we can simply define effective equation of state (EoS) parameter  $w_{eff}$ as:
\begin{equation}
w_{eff}=\sum \frac{p_i}{\rho_i}=\frac{2}{3}(1+q).
\end{equation}
We observe that  $q > -1$ if the expansion of the Universe is decelerating ,while we need $q<-1$ if the cosmic expansion is accelerating.

However, we have still  several models to explain DE and it was suggested that all models can be classified 
 using a set of cosmological parameters called the statefinder pair
$\left\{r, s\right\}$ given by \cite{rs}:
\begin{equation}\label{state8.1}
r\equiv\frac{\stackrel{...}a}{aH^3},~~~~~~~~~~s\equiv\frac{r-1}{3(q-1/2)},
\end{equation}
As usual $(a,H)$ denote the scale factor and the Hubble
parameter.
In any type of DE, a trajectory in $(r,s)$ plane classifies the model's behavior \cite{Albert1,Albert2}. For the LCDM model, we have $s=0$ and $r=1$. As a result, the evolution of the universe at late time could be relevant, i.e., a contribution from matter is negligible. In this situation, the universe expands due to the cosmological constant only and the scale factor grows exponentially with time.

%%%%%%%%%%%%%%%%%%%%%%%%%%%%%%
\section{Concluding remarks}
\label{s6}
\begin{figure}
	\includegraphics[width=6.0cm]{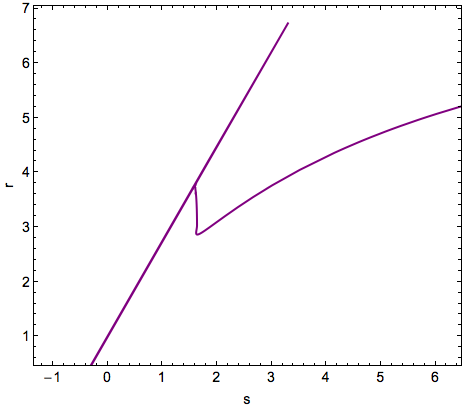}
	\caption{The plot shows the state finder parameters $(s,r)$ for Class A.}\label{qrsA}
\end{figure}

\begin{figure}
	\includegraphics[width=8.0cm]{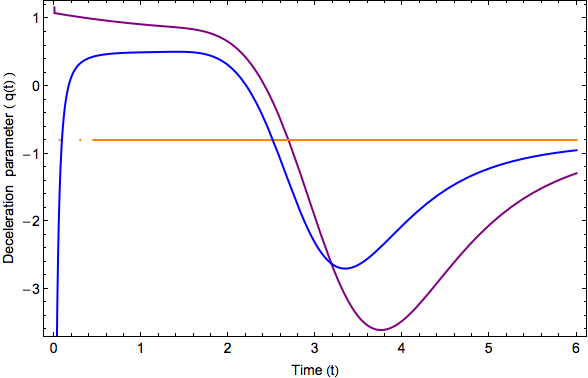}
	\caption{The plot shows a deceleration parameter $q$ for Class A (purple), Class B (orange) and Class C (blue).}\label{dece}
\end{figure}

\begin{figure}
	\includegraphics[width=2.5cm]{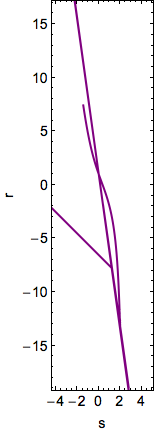}
		\includegraphics[width=3.15cm]{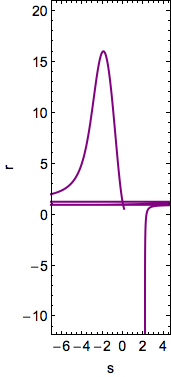}
	\caption{State finder parameters $(s,r)$ for Class B (left-panel) and for Class C (right-panel).} \label{rsss}
\end{figure}

%%%%%%%%%%%%%%%%%%%%%%%%%%%%%
In this work, we considered a formal framework of nonlocal $f(T)$ theory of gravity and examined the nonlocal $f(T)$ theory through the Noether symmetry approach. Here we derived the Noether equations of the nonlocal $f(T)$ theory in flat FLRW universe. We analysed the dynamics of the field in the a nonlocal $f(T)$ gravity using the Noether sysmetry technique. We summarize our results as follows. For class A solutions, we integrate the EOMs using a set of initial conditions (ICs) such that $a(0)= 1, \phi(0)= 2, \xi(0)= 0.5, \dot\xi(0) =0.7, \dot\phi(0)= 1$ and $\dot a(0)=0.6$. Note that these ICs are arbitrarily chosen and the bahavior of fields are independent from these ICs. In Fig.(\ref{dece}), we displays a deceleration parameter $q$ which shows that there exists a transition from a deceleration phase $q>-1$ to an acceleration one $q<-1$. From  Fig.(\ref{qrsA}), we display the statefinder parameters ($s,r$) and discover that our model is compatible the LCDM model for which a point $(0,1)$ is satisfied.
For class B solutions, we integrate the EoMs using a set of initial conditions  such that $a(0)= 0.3, \phi(0)=0.2, \xi(0)= 0.5, \dot\xi(0) =0.5,\dot\phi(0)= -0.5$ and $\dot a(0)=0.5$. Notice that there exist a transition from a deceleration phase to an acceleration phase at small $t$. However, the deceleration parameter $q(t)$ deduces an deceleration phase at late times displayed in Fig.(\ref{dece}). Likewise, it is hard in this class for obtaining $s=0$ and $r=1$ illustrated on the left-hand side of Fig.(\ref{rsss}).

We further continue our investigations for class C solutions by specifying initial conditions: $a(0)= 0.3, \phi(0)=0.2, \xi(0)= 0.5, \dot\xi(0) =0.5, \dot\phi(0)= -0.5$ and $\dot a(0)=0.5$. We demonstrate by considering the statefinder parameters illustrated on the right-panel of Fig.(\ref{rsss}) that the model can be compatible with the LCDM with $s=0$ and $r=1$. Similar to the class A, we have in this class another phase transition from a deceleration to an acceleration as well as an accelerating expansion resulting from $q<-1$. Note that in order to describe the quantum state of the universe, we may in principle examine a quantum version of the present model by following Refs.\cite{DeWitt:1967yk,Hartle:1983ai} based upon on the Wheeler-DeWitt (WDW) equation.

Moreover, in nonlocal GR models, it has recently proved that the Ricci-flat vacuum exact solutions where $R_{\mu\nu}=0$ are stable under linear perturbations. However, this local stability is only valid for  a class of weakly nonlocal gravitational theories. The present work can be linked to the stability of the black holes's point of view. There is no any simple equivalent class of black hole solutions in our nonlocal teleparallel gravity models. If we ignore the higher order (mainly quadratic) terms in the nonlocal action and focus only on the action when it is constructed from linear $T$, this model in the Einstein frame will be equivalent to the teleparallel gravity plus scalar fields terms. In order to pass from one frame to another (here from conformal frame to the Einstein frame), we need to be very careful to address physical quantities, e.g. in generalized teleparallel gravity, see the local form of our model \cite{Yang:2010ji}. It has been proved that scalar hairy black holes in Einstein gravity are unstable under small perturbations, specially in the AdS backgrounds and they undergo a phase transition \cite{Ganchev:2017uuo}. Consequently we can mimic in this nonlocal teleparallel that the same argument holds. In other words, scalar hairy black holes in nonlocal teleparallel gravity are unstable. However the general proof lies beyond the scope of the present manuscript. Here it is left for a future investigation.

%%%%%%%%%%%%%%%%%%%%%%%%%%%%%%
\section*{Acknowledgement}
%%%%%%%%%%%%%%%%%%%%%%%%%%%%%
We thank the anonymous referee for intuitive comments and thorough criticism on our manuscript.

\end{document}